\documentclass[dvipdfmx]{emulateapj-rtx4}

\usepackage{amssymb}
\usepackage{amsmath}
\usepackage{color}
\usepackage{ulem}
\usepackage{bm}
\usepackage{multirow}

\newcommand{\ltsim}{\protect\raisebox{-0.5ex}{$\:\stackrel{\textstyle <}{\sim}\:$}}

\slugcomment{accepted to the Astrophysical Journal}

\shorttitle{The critical mass ratio for violent merger scenario}
\shortauthors{Sato et al.}

\begin{document}

\title{The critical mass ratio of double white dwarf binaries
for violent merger-induced Type Ia supernova explosions}

\author{Yushi Sato$^{1,2}$\altaffilmark{*}, Naohito Nakasato$^{3}$, Ataru Tanikawa$^{2,4}$,
Ken'ichi Nomoto$^{5}$\altaffilmark{**},
Keiichi Maeda$^{6,5}$, Izumi Hachisu$^{2}$}
\affil{$^{1}$Department of Astronomy, Graduate School of Science, The University of Tokyo, 7-3-1 Hongo, 
Bunkyo-ku, Tokyo 113-0033, Japan\\
$^{2}$Department of Earth Science and Astronomy, College of Arts and Sciences, The University of Tokyo,
 3-8-1 Komaba, Meguro-ku, Tokyo 153-8902, Japan\\
$^{3}$Department of Computer Science and Engineering, University of Aizu, Tsuruga Ikki-machi Aizu-Wakamatsu, 
Fukushima 965-8580, Japan\\
$^{4}$RIKEN Advanced Institute for Computational Science, 7-1-26 Minatojima-minami-machi, Chuo-ku, 
Kobe, Hyogo 650-0047, Japan\\
$^{5}$Kavli Institute for the Physics and Mathematics of the Universe (WPI), The University of Tokyo, 
5-1-5 Kashiwanoha, Kashiwa, Chiba 277-8583, Japan\\ 
$^{6}$Department of Astronomy, Kyoto University, Kitashirakawa-Oiwake-cho, Sakyo-ku, Kyoto 606-8502, 
Japan}

\altaffiltext{*}{sato@ea.c.u-tokyo.ac.jp}
\altaffiltext{**}{Hamamatsu Professor}

\begin{abstract}
Mergers of carbon-oxygen (CO) white dwarfs (WDs) are considered as one of 
the potential progenitors of type Ia supernovae (SNe~Ia).
Recent hydrodynamical simulations showed that the less massive (secondary)
WD violently accretes onto the more massive (primary) one,
carbon detonation occurs, the detonation wave propagates
through the primary, and the primary finally explodes as
a sub-Chandrasekhar mass SN Ia.
Such an explosion mechanism is called the violent merger scenario.
Based on the smoothed particle hydrodynamics (SPH)
simulations of merging CO WDs,
we derived more stringent critical mass ratio ($q_{\rm cr}$) leading to
the violent merger scenario than the previous results.
We conclude that this difference mainly comes from the differences
in the initial condition, synchronously spinning of WDs or not.
Using our new results, we estimated the brightness
distribution of SNe Ia in the violent merger scenario 
and compared it with previous studies.
We found that our new $q_{\rm cr}$ does not
significantly affect the brightness distribution.
We present the direct outcome immediately following CO WD mergers
for various primary masses and mass ratios.
We also discussed the final fate of the central system of 
the bipolar planetary nebula Henize 2-428,
which was recently suggested to be
a double CO WD system whose total mass exceeds
the Chandrasekhar-limiting mass,
merging within the Hubble time.
Even considering the uncertainties in the proposed binary parameters,
we concluded that the final fate of this system
is almost certainly a sub-Chandrasekhar mass SN~Ia
in the violent merger scenario.
\end{abstract}


\section{Introduction}
\label{introduction}
SNe Ia have been considered as 
thermonuclear explosions of a CO WD.
They play important roles in astronomy as
a standard candle to determine the cosmological parameters
\citep[e.g.,][]{riess98, perlmutter99} and as major sources of 
iron group elements \citep[e.g.,][]{kobayashi98}.
However, their progenitor systems and explosion mechanisms
are still uncertain \citep[e.g.,][]{hillebrandt00, maoz14}.
A merger of CO WDs has been considered as one of 
the potential progenitors of SNe Ia, which is called
the double degenerate (DD) scenario \citep{iben84, webbink84}.
A binary of CO WDs loses its angular momentum 
by emitting gravitational waves and finally merges.
When the total mass exceeds
the critical mass\citep[$M_{\rm ig}=1.38~M_{\odot}$,][]{nomoto84}
for the carbon ignition in the center,
such a system has been proposed to
give rise to an SN Ia explosion.
Some observational studies reported that 
neither surviving companions 
\citep[e.g.,][]{gonzalez12, kerzendorf12, schaefer12}
nor signatures of them \citep[e.g.,][]{brown12, foley12, olling15}
were detected in some well-studied SN~Ia remnants and
nearby SNe~Ia,
but recently there are several examples
which possibly show these signatures as
well both in SNe with special properties
\citep{mccully14, cao15} and in a normal
SN Ia \citep{marion15}.
These studies suggest that at least a part of SNe Ia would
originate from either the DD scenario or the $J$(angular
momentum)-loss induced ``delayed''
carbon ignition in the single degenerate (SD)
scenario \citep[e.g.,][]{benvenuto15}.

After \citet{benz90} firstly performed three dimensional
(3D) hydrodynamical simulations of CO WD mergers with 
their SPH code
\citep[e.g.,][]{gingold82, monaghan05, rosswog09},
similar merger simulations have been performed by various groups
\citep[e.g.,][]{rasio95, segretain97, guerrero04, 
dsouza06, motl07, yoon07, loren09, fryer10,
pakmor10, pakmor11, pakmor12a, pakmor12b,
dan11, dan12, dan14, raskin12, raskin14, zhu13,
moll14, kashyap15, sato15, tanikawa15}.
These studies showed that a CO WD merger could lead to
an SN Ia if several conditions are satisfied.
\citet{pakmor10, pakmor11, pakmor12a, pakmor12b} 
simulated mergers of CO WDs, both of the binary members 
have masses ${\sim}~1~M_{\odot}$, with their SPH code.
They found that the secondary WD violently accreted
onto the primary and carbon detonation would occur
during the merger.
They showed that detonation wave propagated through
the primary and the primary finally exploded as an SN Ia.
They called this explosion mechanism 
``the violent merger scenario.''
As the mass ratio $q~{\equiv}~M_{2}/M_{1}$,
where $M_1$ is the primary mass and
$M_2$ is the secondary mass,
approaches unity, the mass accretion becomes more violent
and the carbon detonation occurs more easily.
Therefore, the mass ratio is an important parameter
for the violent merger scenario.

\citet{pakmor11} investigated the critical mass ratio
$q_{\rm cr}$ above which the violent merger scenario
is realized,
and found that $q_{\rm cr}\sim~0.8$ for $M_{1}~=~0.9~M_{\odot}$.
\citet{dan12} studied mergers of WDs over
a wide range of masses, i.e., 
$0.2~{\sim}~1.2~M_{\odot}$.
They found that dynamical carbon burning
did not occur in all their models.
However, their numerical resolution was low
(${\sim}~20,000$ particles per star).
\citet{sato15} performed SPH simulations of CO WD mergers
over a wide range of masses $(0.5~{\sim}~1.1~M_{\odot})$
and with a resolution of 500,000 particles per $M_{\odot}$,
i.e., $500,000~M_{\odot}^{-1}$.
They investigated whether or not dynamical carbon burning would occur
during the merger.
\citet{pakmor12b} and \citet{sato15} showed that 
the numerical resolution is another important
factor in studying the violent merger scenario, and 
\citet{sato15} suggested that it should be 
${\ge}~500,000~M_\sun^{-1}$
\citep[see also ][]{tanikawa15}.

Assuming critical mass ratios for the violent merger scenario
based on the result of \citet{pakmor11}, \citet{ruiter13}
estimated the brightness distribution of SNe~Ia and suggested that
it could be qualitatively consistent with what is observed \citep{li11}.
Because the critical mass ratio could have influence on the distribution
of SN~Ia brightness,
detailed investigation of the critical mass ratio is
required to verify the violent merger scenario.

Recently, \citet{santander15} reported that 
the central system of
the bipolar planetary nebula (PN) Henize 2-428
might be a super-Chandrasekhar DD pair.
They estimated the combined mass as ${\sim}~1.8~M_{\odot}$
and the mass ratio as ${\sim}~1$.
Although there are some negative arguments against
their conclusion \citep[e.g.,][]{frew15, garcia15},
Henize 2-428 is a candidate progenitor 
for the violent merger scenario,
if the interpretation by \citet{santander15} is correct.
While \citet{sato15} have already simulated a set of primary and
secondary masses which cover a plausible parameter space for Henize 2-428,
their mass grids were too coarse.
In this paper, we perform some additional (much finer grids of the masses)
SPH simulations of CO WD mergers to accurately obtain
the critical mass ratio for the violent merger scenario.
We try to derive the relation between $M_1$ and
$q_{\rm cr}$, and compare it with previous studies.
We also use our results to discuss
the final fate of Henize 2-428.

This paper is organized as follows. 
In Section \ref{method}, we briefly describe
our numerical method.
Our results are shown in Section \ref{results}
and we discuss them in Section \ref{discussion}.
We summarize our findings in Section \ref{summary}.

\begin{figure}
  \begin{center}
        \includegraphics[width=8.0cm, angle=0]{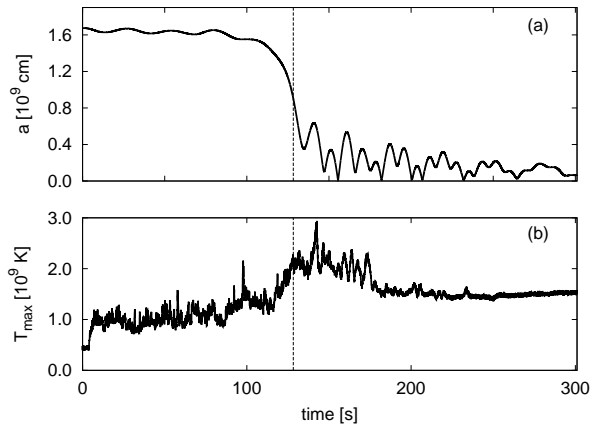}
  \end{center}
  \vspace{5pt}
 \caption{Time evolutions of (a) the orbital separation
 and (b) the maximum temperature in the case of
 $1.1+0.9~M_\sun$.
 Vertical dashed lines show the time when the first
 detonating particle appears.
}
 \label{fig1}
\end{figure}

Figure \ref{fig3} shows the density and
temperature of the particle having smallest
$\tau_{\rm CC}/\tau_{\rm dyn}$
ratio in each merger simulation, where 
$\tau_{\rm CC}$ and $\tau_{\rm dyn}$ are carbon burning
and dynamical timescales, respectively, as defined below.
Colors and shapes of symbols indicate
the total mass, $M_{\rm tot}=M_1+M_2$,
and primary mass, $M_1$, respectively.
A black solid line shows the line of
$\tau_{\rm CC}/\tau_{\rm dyn}=1$.

\section{Numerical Methods}
\label{method}
Our SPH simulations of CO WD mergers are essentially
the same as those of \citet{sato15}.
We used the OcTree On OpenCL (OTOO) code,
which was developed for various particle simulations 
for astrophysical fluid phenomena \citep{nakasato12}.
The equation of state (EOS) is that of \citet{timmes00}.
Initial setups are also the same as those in \citet{sato15}, 
which refer to \citet{rasio95} and \citet{dan11}.
Our CO WD models have an uniform composition 
of $50\%$ carbon and $50\%$ oxygen in mass.
\citet{sato15} simulated mergers
of CO WDs whose masses were 0.5, 0.6, 0.7,
0.8, 0.9, 1.0, and $1.1~M_{\odot}$.

In this paper, we performed additional simulations of CO WD
mergers to accurately determine the critical mass ratio 
$q_{\rm cr}$ to realize the violent merger
scenario.  We add $M_{2}~=$ 0.725, 0.75, 0.775, 0.825, 0.85, 
and $0.875~M_{\odot}$ for each $M_{1}~=$ 0.75, 0.8, 0.9, 1.0, 
and $1.1~M_{\odot}$, because $q_{\rm cr}$ seems to be in this mass
range \citep[see Figure 5 of][]{sato15}.
The numerical resolution is $500~k$
per solar mass $(k~{\equiv}~1,024)$.
Our SPH scheme derives the temperature of a particle from the density
and internal energy through the EOS.
The estimations of density and internal energy
have some numerical noises, and the derived temperature
is also fluctuated \citep{dan12, sato15, tanikawa15}.

Because nuclear burning rate is sensitive to the temperature,
it might strongly enhance the fluctuated temperature.
Therefore, we firstly performed our simulations
without nuclear reactions, as in our previous work \citep{sato15},
and determined the critical mass ratio for the violent merger scenario.
Then, we added simulations including nuclear reactions 
in order to confirm that the inclusion of nuclear burning
in our SPH simulation increases the temperature high enough 
to really satisfy the detonation condition for
the models above the critical mass ratio 
(see Section \ref{nuclear_burning}).

\section{Results}
\label{results}
\begin{figure}
  \begin{center}

        \includegraphics[width=7.0cm, angle=0]{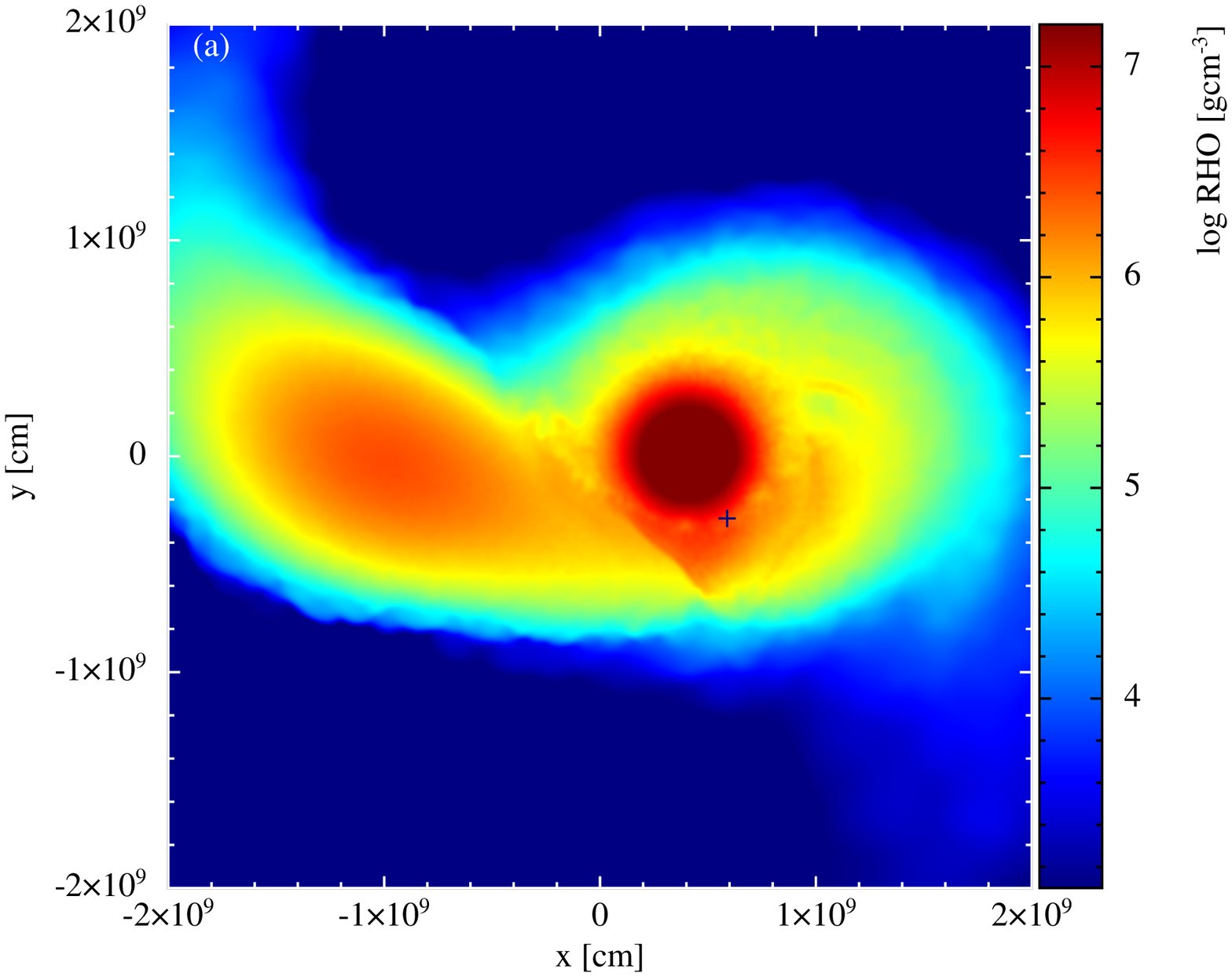}

        \includegraphics[width=7.0cm, angle=0]{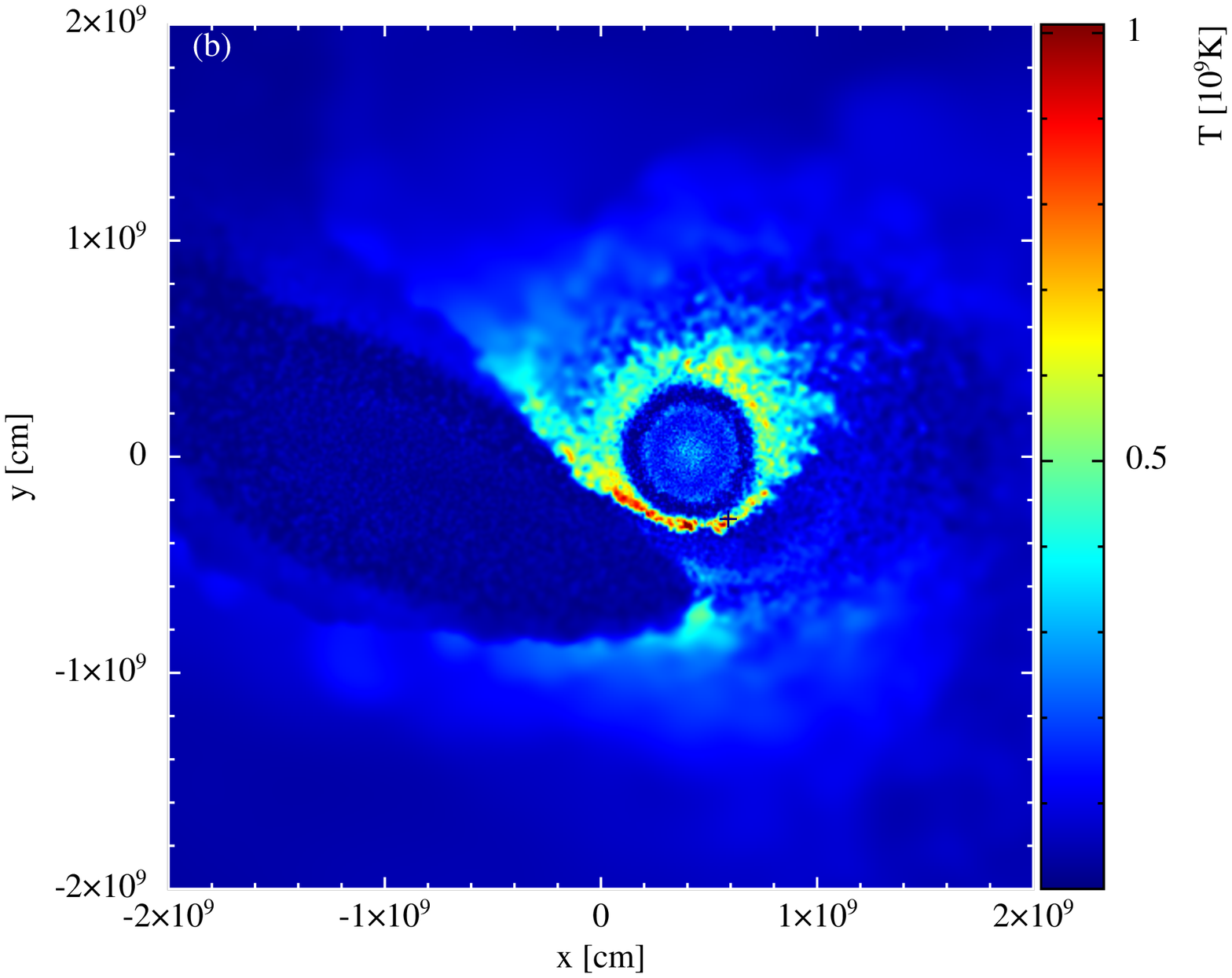}

  \end{center}
  \vspace{5pt}
 \caption{
 (a) Density and (b) temperature profiles on the equatorial plane at
 the time of dashed line in Figure \ref{fig1}.
 A black cross indicates the first detonating
 particle.
}
 \label{fig2}
\end{figure}

Figure \ref{fig1}  
shows time evolutions of the orbital separation
and the maximum temperature in the case of
$1.1+0.9~M_\sun$.
They indicate that the temperature increases drastically
when the secondary is completely disrupted and accretes
onto the primary, which is consistent with
previous studies.
Black vertical dashed lines show the time when
our detonation condition, which is described
below in Section \ref{condition}, is first fulfilled.
Figure \ref{fig2}  
depicts the density and temperature
profiles of the equatorial plane at
the time which is indicated by the
black dashed lines in Figure \ref{fig1}.
The profile and morphology are consistent with the similar cases
in \citet{pakmor12a,pakmor12b}.

\subsection{Conditions for Violent Merger-Induced Explosion}
\label{condition}
For the violent merger scenario, it is crucial whether or not 
carbon detonation occurs during a merger.
Although \citet{pakmor11} used the results of 
\citet{seitenzahl09} as the detonation condition,
the treatment is still controversial.
In this paper, we judge it from the condition to
trigger the dynamical carbon burning.
We obtain the dynamical timescale \citep[e.g.,][]{nomoto82},
\begin{equation}
\tau_{\rm dyn} = \frac{1}{\sqrt{24{\pi}G\rho}},
\end{equation}
and the carbon burning timescale,
\begin{equation}
\tau_{\rm CC} = \frac{{C_P T}}{ \epsilon_{\rm CC}},
\end{equation}
for each SPH particle.
Here, $\rho$ and $T$ are particle's density and temperature,
respectively, $C_P$ is the specific heat at constant pressure
of a particle, $G$ is the gravitational constant, and
$\epsilon_{\rm CC}$ is the energy generation rate of
carbon burning \citep[see Equation (6) of][]{sato15}.
When there are any particles that satisfy
$\tau_{\rm CC} < \tau_{\rm dyn}$,
carbon ignites dynamically.
We regard this as the detonation condition.

If $\tau_{\rm CC} < \tau_{\rm dyn}$,
the temperature increases more rapidly than decreases by
adiabatic expansion and accelerates carbon burning.
The temperature would increase high enough to satisfy
the detonation condition, such as the results of \citet{seitenzahl09}.
Thus, we consider that the condition of the dynamical carbon
burning ($\tau_{\rm CC} < \tau_{\rm dyn}$)
is a plausible detonation condition in the case without
nuclear reactions.  We confirm this later by including 
carbon burning in our SPH code in Section \ref{nuclear_burning}.

As described in Section \ref{method},
temperature is possibly affected by
numerical noises in our SPH simulations.
In order to minimize the influence of temperature
noise in judging the condition for carbon detonation,
we examine whether the particle continuously satisfies
the above condition at least for a dynamical timescale
($\tau_{\rm dyn}~{\sim}~0.4$~s at $\rho\sim10^{6}$~g~cm$^{-3}$). 
If so, we assume that dynamical carbon burning is definitely
initiated and detonation occurs.

\begin{figure}
  \begin{center}
    
        \includegraphics[width=8.0cm, angle=0]{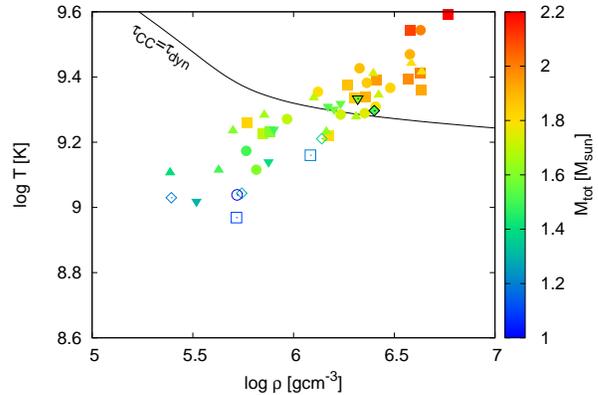}

  \end{center}
  \vspace{5pt}
 \caption{Density and temperature of a particle having
  smallest $\tau_{\rm CC}/\tau_{\rm dyn}$ ratio
  for each CO WD merger model.
  Colors of symbols indicate the total mass of the system.
  Shapes of symbols do the mass of the primary. 
  Filled squares are the $1.1~M_{\odot}$ primary, filled circles 
  $1.0~M_{\odot}$, filled triangles $0.9 ~M_{\odot}$, filled inverted
  triangles $0.8~M_{\odot}$, filled diamonds $0.75~M_{\odot}$, 
  open diamonds $0.7~M_{\odot}$, open squares $0.6~M_{\odot}$, 
  open circles $0.5~M_{\odot}$.
  Symbols surrounded by black frames indicate $0.8+0.8~M_{\odot}$
  (filled inverted triangle) and $0.75+0.75~M_{\odot}$ (filled diamond), 
  respectively.
  A black solid line indicates $\tau_{\rm CC}~=~\tau_{\rm dyn}$.}
 \label{fig3}
\end{figure}

For example, mergers of $0.8+0.8~M_{\odot}$
and $0.75+0.75~M_{\odot}$, which are surrounded by
black frames in Figure \ref{fig3}, 
both satisfy $\tau_{\rm CC} < \tau_{\rm dyn}$.
However, in the case of $0.75+0.75~M_{\odot}$,
there are no particles which continuously satisfy the above condition
for a dynamical timescale.   
We thus regard that detonation does
not occur in the merger of $0.75+0.75~M_{\odot}$.
In the case of $0.8+0.8~M_{\odot}$, on the other hand,
three particles keep $\tau_{\rm CC} < \tau_{\rm dyn}$
for longer than $\tau_{\rm dyn}$.
Therefore, we regard that carbon detonation does occur
in the case of $0.8+0.8~M_{\odot}$ and eventually the system leads to
an SN Ia in the violent merger scenario.

\begin{figure}
  \begin{center}
    
        \includegraphics[width=8.0cm, angle=0]{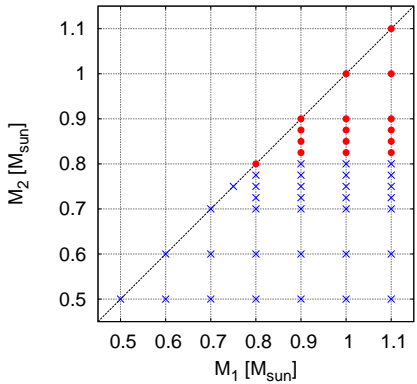}

  \end{center}
  \vspace{5pt}
 \caption{Filled circles indicate the mass combinations which
          satisfy our condition for carbon detonation
          while crosses do ones which do not satisfy the condition.
          A black dashed line is where the mass ratio equals
          to unity.}
 \label{fig4}
\end{figure}

\subsection{Critical Mass Ratio for Violent Merger Scenario}
\label{critical}
We plot our numerical results of violent merger scenario
in Figure \ref{fig4}.  
Here, the filled circles denote models
which satisfy our condition for carbon detonation while the crosses
do models which do not.
The black dashed line indicates $M_1~=~M_2$.
Figure \ref{fig4} shows that carbon detonation occurs only in
the models whose primary and secondary masses are more
massive than ${\sim}~0.8~M_{\odot}$, which is consistent
with the results of \citet{sato15}.

We derived approximate formulae of the critical mass ratio,
$q_{\rm cr}$, as a function of the primary mass, $M_1$,
and plot them in Figure \ref{fig5},  
both of which are on logarithmic scales. 
The red solid line is for an upper bound,
while the blue solid line is for a lower bound.
They are represented by
\begin{equation}
q_{\rm cr}=0.82\left(\frac{M_1}{M_{\odot}}\right)^{-0.91}~({\rm upper~bound}),\\
\label{upper_bound}
\end{equation}
\begin{equation}
q_{\rm cr}=0.80\left(\frac{M_1}{M_{\odot}}\right)^{-0.84}~({\rm lower~bound}),
\label{lower_bound}
\end{equation}
respectively.
Considering the lowest mass ratios in which the model satisfies
our condition for carbon detonation, we derive Equation
(\ref{upper_bound}).  On the other hand, we derive Equation
(\ref{lower_bound}) from the highest mass ratios in which
the model does not satisfy our condition for carbon detonation.

If a sizable helium layer exists on the CO WD,
helium detonation could occur during merger
and induce carbon detonation
\citep[e.g.][]{guillochon10,dan12,raskin12,pakmor13,tanikawa15}.
In this helium-ignited violent merger scenario,
$q_{\rm cr}$ could be smaller than Equations (\ref{upper_bound})
and (\ref{lower_bound}).
But such a study needs much finer resolution
because the occurrence of helium detonation depends
strongly on the mass of the helium layer \citep[e.g.,][]{tanikawa15}.
We leave such a systematic study to our future work.

\section{Discussion}
\label{discussion}

\subsection{Comparison with previous studies}
\label{comparison}
\citet{pakmor11} obtained $q_{\rm cr}~{\sim}~0.8$ for
$M_1~=~0.9~M_{\odot}$.
Our results show a different value of $q_{\rm cr}~{\sim}~0.9$
for $M_1~=~0.9~M_{\odot}$.
This discrepancy could arise from the differences in 
(1) the initial conditions,
(2) numerical resolution, and 
(3) inclusion of nuclear burning.
We discuss these three effects in this order.

\begin{figure}
 \begin{center}
   
        \includegraphics[width=8.0cm, angle=0]{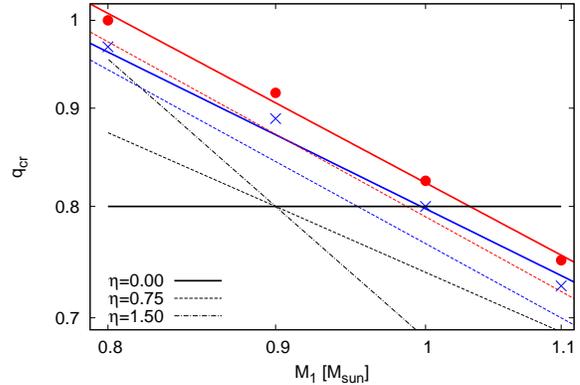}

 \end{center}
 \vspace{5pt}
 \caption{The critical mass ratio for the violent merger scenario
   as a function of the primary mass.
   A red solid line denotes an upper bound represented by Equation
   (\ref{upper_bound}),
   while a blue solid line shows a lower bound represented by
   Equation (\ref{lower_bound}).
   {We also present the results including carbon burning.
   Red and blue dashed lines correspond to the upper and lower bounds
   for the violent merger scenario in the case with carbon burning, 
   respectively.}
   The critical mass ratios adopted by \citet{ruiter13}
   are also depicted by a black solid, dashed, and
   dash-dotted lines.
   See text for more details.}
 \label{fig5}
\end{figure}

\citet{pakmor11} used non-spinning WDs
and set them at the initial separation where the 
Roche lobe overflow (RLOF) just started in the
sense of Eggleton's (1983) approximation.
On the other hand, we assume that the WDs
spin synchronously with the orbital motion,
and set them at the initial separation large enough to avoid the RLOF.
We decrease the separation
until the RLOF starts \citep{rasio95, dan11, sato15}.
For the initial condition of \citet{pakmor11},
as mentioned in \citet{dan11},
the secondary accretes onto the primary
more violently than for the case of synchronously spinning WDs.
As a result, carbon detonation would occur easily.
We examined the merger of $0.9+0.75~M_{\odot}$ WDs
in the similar initial condition to that of \citet{pakmor11}.
We thereby found that there exist 16 particles which satisfy
the condition for carbon detonation in this case,
while there is no such a particle for our
initial condition (see Figure \ref{fig3}).
Although no definite conclusion has been reached yet,
several numerical studies suggest that 
compact WD binaries would reach synchronization
before their merging due to angular momentum
dissipation by tidally excited gravity waves
\citep[e.g.][ and references therein]{fuller14}.

\subsection{Numerical resolution}
\label{numerical_resolution}
\begin{figure}
  \begin{center}
    
        \includegraphics[width=8.0cm, angle=0]{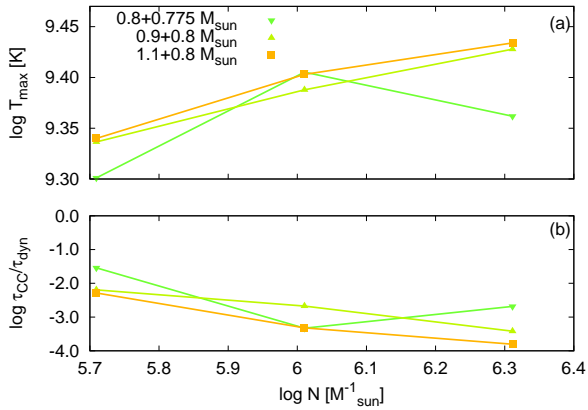}

  \end{center}
  \vspace{5pt}
 \caption{Dependence of (a) the maximum temperature
 and (b) the minimum ${\tau}_{\rm CC}/{\tau}_{\rm dyn}$
 ratio on the numerical resolution.
 The horizontal axis is the number of SPH particles 
 per solar mass.
 The shapes and colors of symbols have the same meanings
 as those in Figure \ref{fig3}.}
 \label{fig6}
\end{figure}

Numerical resolution might also cause the difference.
\citet{sato15} reported that the maximum temperature and
the minimum ${\tau}_{\rm CC}/{\tau}_{\rm dyn}$ ratio
did not converge in the range of $10k~{\sim}~500k~M^{-1}_{\odot}$.
This is because smaller hot regions can be resolved
as the numerical resolution becomes higher.
Several previous studies also indicated that
numerical resolution would affect their results
\citep[e.g.,][]{pakmor11, pakmor12b}. 
In this paper, we performed several simulations with higher
numerical resolutions ($1000k~M^{-1}_{\odot}$, $2000k~M^{-1}_{\odot}$)
than our standard ones ($500k~M^{-1}_{\odot}$),
for $0.8+0.775~M_{\odot}$, $0.9+0.8~M_{\odot}$,
and $1.1+0.8~M_{\odot}$ WDs.
Our numerical results are summarized in Figure \ref{fig6}.
The maximum temperature still increases
and the minimum ${\tau}_{\rm CC}/{\tau}_{\rm dyn}$ ratio decreases
except for $0.8+0.775~M_\sun$ as the numerical resolution increases.
Similar trends were also reported in \citet{pakmor12b},
\citet{sato15}, and \citet{tanikawa15}.
These trends indicate that our numerical results have not been fully converged yet.
Therefore, our critical mass ratio could be slightly
lower than the red solid line in Figure \ref{fig5}.

\subsection{Effect of nuclear burning}
\label{nuclear_burning}
Inclusion of nuclear reactions could also make
the difference between theirs and ours.
\citet{pakmor11} found that particles with high temperature
(${\geq}~2{\times}10^{9}$~K) did not exist in the case of
$0.9+0.81~M_{\odot}$ without nuclear reactions.
On the other hand, they existed in the case
with nuclear reactions.

In order to evaluate the effect of nuclear energy release,
we performed additional simulations including
nuclear burning for some models.
In these simulations, we include only
carbon burning, i.e.,
${}^{12}{\rm C}~+~{}^{12}{\rm C}$ reaction,
to avoid large computational cost
for solving a reaction network.
The formulation of the reaction rate is the same as
Equation (6) of our previous work \citep{sato15}.
Because the initiation of carbon detonation
is judged by this reaction,
such a simple treatment of nuclear reactions
would be sufficient for our purpose.
Electron screening effect is negligibly small
in the region of $\rho=10^{6}$--$10^{7}$~g~cm$^{-3}$
and $T > 2\times 10^9$~K, where carbon detonation would occur,
so we do not include this effect.

\begin{figure}
 \begin{center}
   
        \includegraphics[width=8.0cm, angle=0]{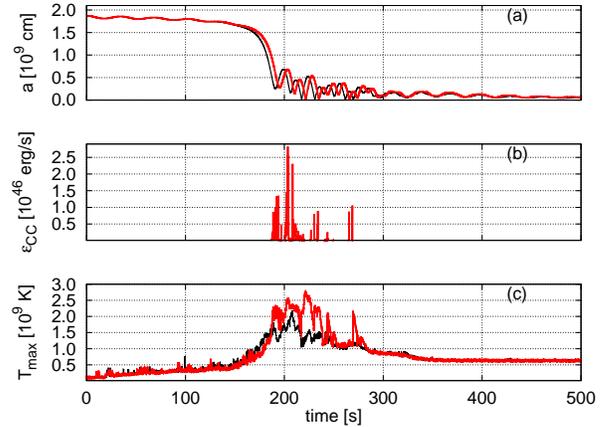}

 \end{center}
 \vspace{10pt}
 \caption{Merger of $0.9+0.8~M_{\sun}$ WDs.
 Red lines indicate the case
 with carbon burning,
 while black ones do the case without.
(a) The orbital evolution,
(b) the energy release rate  of carbon burning, and
(c) the time evolution of maximum temperature.
}
 \label{fig7}
\end{figure}

\begin{figure}
 \begin{center}
   
        \includegraphics[width=8.0cm, angle=0]{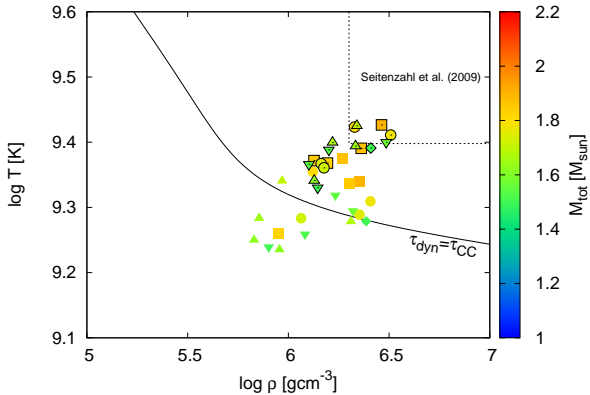}

 \end{center}
 \vspace{10pt}
 \caption{Same as Figure \ref{fig3}, but for the case with carbon burning.
The symbols surrounded by black frames are the case with carbon
burning while not surrounded are the case without.
Black dashed line is the demarcation of detonation condition
by \citet{seitenzahl09}.
}
 \label{fig8}
\end{figure}

Figure \ref{fig7} shows the comparison between simulations
with and without carbon burning for the case of $0.9~+~0.8~M_{\odot}$.
The red lines indicate the case with carbon burning,
while the black ones do the case without.
Although the orbital evolution (Figure \ref{fig7}(a))
is almost the same, the maximum temperature
(Figure \ref{fig7}(c)) realized in the case with carbon burning
is higher than the case without during the merger phase.
This difference comes from the nuclear energy release (Figure \ref{fig7}(b)).
We confirmed that the inclusion of carbon burning
increase the maximum temperature.
Figure \ref{fig8} summarizes the density and temperature
of a particle which has the smallest 
${\tau}_{\rm CC}/{\tau}_{\rm dyn}$ ratio for all the cases 
we recalculated including nuclear burning.
The symbols surrounded by black frames are
the results of the cases with carbon burning.

\begin{figure}
 \begin{center}
   
        \includegraphics[width=8.0cm, angle=0]{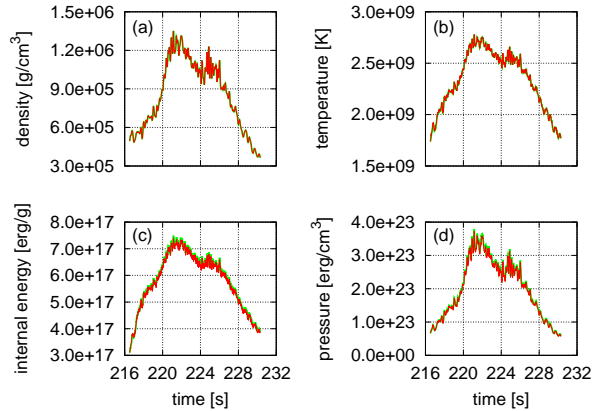}

 \end{center}
 \vspace{10pt}
 \caption{
Post-processing analysis
 in the case of $0.9~+~0.8~{\rm M_{\odot}}$,
 We have calculated nuclear energy generation} using
 the 13 species $\alpha$-chain reaction network
 \citep{timmes99}.
 Green lines denote the case of only carbon burning while
 red ones do those of post-processing results.
 (a) Density, (b) temperature, (c) specific
 internal energy, and (d) pressure.
 \label{fig9}
\end{figure}

\begin{figure}
 \begin{center}
   
        \includegraphics[width=8.0cm, angle=0]{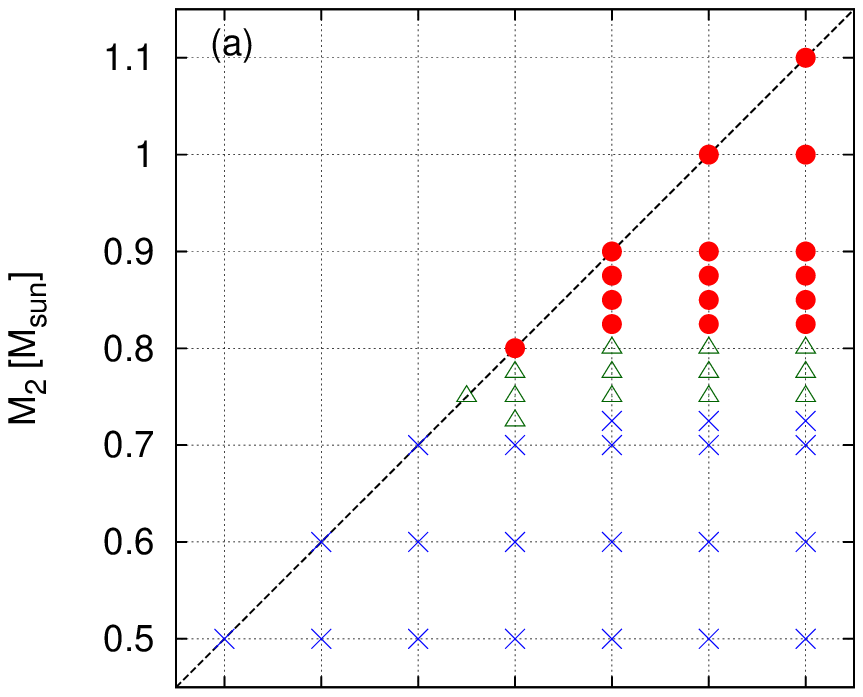}

        \includegraphics[width=8.0cm, angle=0]{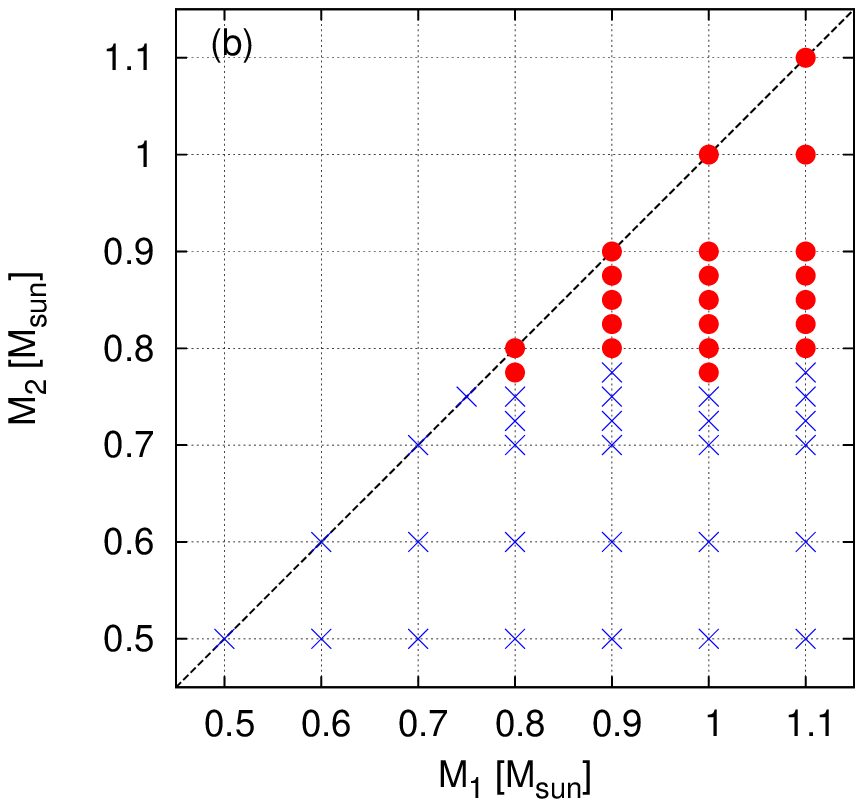}

 \end{center}
 \vspace{10pt}
 \caption{
Same as Figure \ref{fig4}, but for the case with nuclear burning.
Filled circles and crosses have
 the same meanings as those in Figure \ref{fig4}.
 Green triangles indicate the mass combinations
 which satisfy the detonation condition
 with carbon burning although they do not 
 without carbon burning.
 (a) The results based on our detonation condition in Section 
\ref{critical}.
 (b) The results based on the detonation condition of
 \citet{seitenzahl09}.
}
 \label{fig10}
\end{figure}

Since our simulations include only carbon burning,
the effect of nuclear reactions would be underestimated.
To check this effect, we perform post-processing
calculations, using an $\alpha$-chain reaction
network containing 13 species from $^4$He
to $^{56}$Ni \citep{timmes99}.
Figure \ref{fig9} shows the results for
a particle which has the highest temperature
in the case of $0.9~+~0.8~M_{\odot}$.
This calculation was done for a period
in which carbon burning continued.
The red lines indicate the case with
the $\alpha$-chain reaction network,
while green lines do the case with
only carbon burning.
Because we find essentially no difference
between them, our simulations including only
carbon burning are appropriated to estimate the effect
of nuclear burning, and being consistent with
the previous studies using $\alpha$-chain networks
\citep{pakmor10, pakmor11, pakmor12a, raskin12},
at least until the initiation of detonation.
Thus, the inclusion of carbon burning lowers $q_{\rm cr}$
to realize violent merger-induced explosion
if we use the same criterion
as in Section \ref{critical}.
In Figure \ref{fig10}(a), green triangles are
the models which satisfy our detonation condition
with carbon burning, although they do not satisfy
the condition without.

However, our detonation condition in Section \ref{critical}
is considered under the situation without
nuclear burning, so it would be no longer
valid in the case with carbon burning.
Therefore, for comparison with the results
of \citet{pakmor11}, we adopt
the same detonation condition as theirs,
i.e., the detection of particles having
$\rho > 2\times 10^6$~g~cm$^{-3}$
and $T > 2.5\times 10^9$~K \citep{seitenzahl09}.
The results are presented in Figure \ref{fig10}(b).
Although the boundary of Figure \ref{fig10}(b)
is slightly lower than that of Figure \ref{fig4},
the difference between them is small.
Thus, we confirm that the detonation condition
adopted in Section \ref{condition}
is plausible for the case without nuclear reactions.

{The approximated lines of $q_{\rm cr}(M_1)$ in the case with
carbon burning are also presented in Figure \ref{fig5},
where the red (blue) dashed line indicates
an upper (lower) bound.}
The value of $q_{\rm cr}$ with carbon burning is
slightly lower than that without carbon burning.
The lower bound without carbon burning is almost the same as
the upper bound with carbon burning, so we adopt the boundary
of Equation (\ref{lower_bound}).
Then, our $q_{\rm cr}$ is more stringent than that derived in
\citet{pakmor11}.
This difference mainly comes from
the initial condition as discussed in Section \ref{comparison}.

\subsection{Brightness distribution in violent merger scenario}
\label{brightness}
Using the results of \citet{pakmor11}
and the binary population synthesis (BPS) calculation
\citep{ruiter11},
\citet{ruiter13} estimated the brightness distribution of SNe Ia
arising from the violent merger scenario.
They adopted the value of $q_{\rm cr}~=~0.8$ for
$M_1=0.9~M_{\odot}$ from \citet{pakmor11},
and assumed
\begin{equation}
q_{\rm cr}=\min\left[0.8\left(\frac{M_1}{0.9~M_{\odot}}\right)^{-\eta},
1.0\right],
\end{equation}
where they used $\eta~=~0.0,~0.75,~1.5$.
We add these three lines in Figure \ref{fig5}.
Our $q_{\rm cr}(M_1)$ is basically more stringent
than Ruiter et al.'s (2013) assumptions.

In order to examine how our $q_{\rm cr}(M_1)$
would affect the brightness distribution of
the violent merger scenario,
we adopt the same assumptions as in \citet{ruiter13}
except for $q_{\rm cr}(M_1)$.

\begin{figure}
  \begin{center}
    
        \includegraphics[width=8.0cm, angle=0]{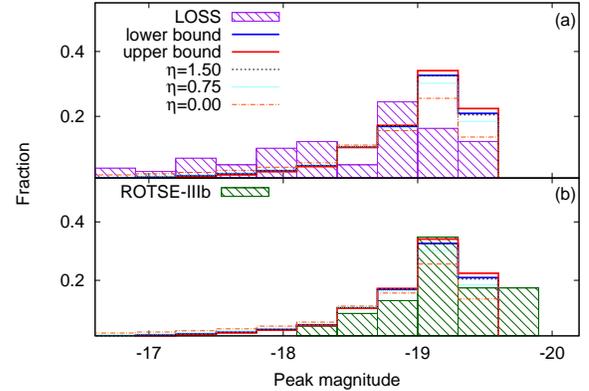}

  \end{center}
  \vspace{10pt}
 \caption{{The SN~Ia brightness distributions
 derived from the present and
 Ruiter et al.'s (2013) results
 are compared with the observations.
 A blue (red) histogram depicts
 the result calculated from the lower (upper)
 bound of our $q_{\rm cr}$.
 Black, cyan, and orange ones
 are derived from hypothetical
 $q_{\rm cr}$ of \citet{ruiter13}.
 A hatched purple} one in panel (a) is a volume-limited
 brightness distribution derived
 by LOSS \citep{li11},
 while a hatched green one in panel (b)
 is that obtained by ROTSE-IIIb
 \citep{quimby12}.
 Magnitudes of models are
 bolometric, while those of
 LOSS and ROTSE-IIIb
 are $R$ band.}
 \label{fig11}
\end{figure}

For the primary at the time of merging, 
we use the same WD mass distribution as \citet{ruiter13}
\citep[see also][]{ruiter11},
although it is highly uncertain
whether the WD can increase its mass
by avoiding the formation of a common envelope
during the very rapid accretion from the He star companion.
We also assume the same $m_{\rm WD}-M_{\rm bol}$
relation of SNe Ia as Ruiter et al.'s (2013)
\citep[see Figure 4 in][]{ruiter13, sim10}
and a flat mass ratio distribution 
of DD systems for simplicity.
Using either our critical mass ratio or theirs,
we calculated the brightness distribution
and compared them each other (Figure \ref{fig11}).
We find that there is no significant
differences between them qualitatively.
This result implies that the $q_{\rm cr}$
of the violent merger-induced explosion
is not so crucial for the brightness
distribution of SNe~Ia, 
as mentioned in \citet{ruiter13}.
Our results are closest to
the ${\eta}~=~1.5$ case among Ruiter et al.'s three cases.
In the both cases of our $q_{\rm cr}$ and 
Ruiter et al.'s $\eta~=~1.5$,
the brightness distribution concentrates around
$-19.0$ mag.
In Figure \ref{fig11}, we add the observational
volume-limited brightness distributions of SNe~Ia.
One in panel (a) is derived by the Lick Observatory Supernova
Search \citep[LOSS, ][]{li11},
and the other in panel (b) is obtained by ROTSE-IIIb  
\citep[see Table 1 in ][]{quimby12}.
The fraction of faint events in our models 
is lower than the observation of LOSS.
This discrepancy might decrease
if we consider the viewing angle effects
\citep[see also discussion in][]{ruiter13},
which might be increasingly important
especially for a large value of $q_{\rm cr}$.
On the other hand, the results of ROTSE-IIIb
are consistent with ours, except most luminous
(${\ltsim}~-19.5$ mag) events.
It should be noted, however, that there are large uncertainties for
the observational brightness distribution
\citep[e.g., ][]{quimby12} and our models are too simple to
be compared with the observational results.
More detailed studies are required to reach
the definitive conclusion.

As the mass ratio is approaching unity,
the primary is more strongly deformed
by the secondary at merging.
As a result, the central density
of the primary becomes lower.
Because the nucleosynthesis in SN~Ia explosions
is so sensitive to the density profile near the center
of the primary in the violent merger scenario,
the amount of $^{56}$Ni decreases
as the deformation of the primary becomes higher
\citep{pakmor10, pakmor11}.
Since our $q_{\rm cr}$ is higher than that
in the previous studies,  
the amount of $^{56}$Ni in the violent merger scenario
would decrease, while this effects are not included
in our brightness estimates.

\subsection{Outcome of mergers and final fate of Henize 2-428}
\label{henize}
\begin{figure}
 \begin{center}
   
        \includegraphics[width=8.0cm, angle=0]{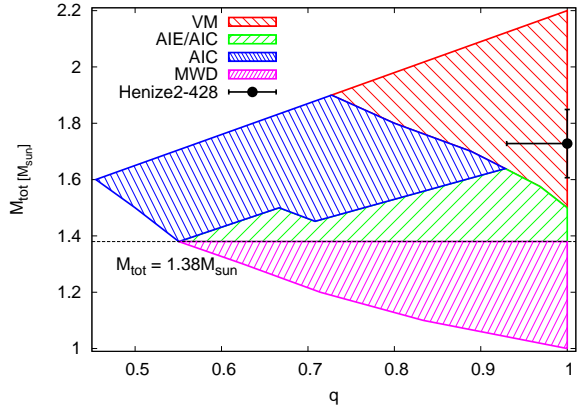}

 \end{center}
 \caption{Final outcomes of our merger simulations
   in the total mass versus mass ratio diagram.
   We adopt Equation (\ref{lower_bound}) as the boundary between the
   VM and AIC or AIE/AIC.
   The boundary between the AIE/AIC and AIC is derived from \citet{sato15}.
   The blanck (white) regions are where the mass of a WD is $<~0.5~M_{\odot}$
   (He WD region) or $>~1.1~M_{\odot}$ (ONeMg WD region).
   He and ONeMg WDs are not the subject of the present study.
   The black dotted line represents 
   $M_{\rm tot}= M_1 + M_2 = M_{\rm ig} = 1.38~M_\odot$.
   If the total mass does not exceed this value,
   the merger remnant becomes a WD without exploding
   as an SN Ia \citep{nomoto84}.
   The black point shows the possible ranges of the mass ratio
   and total mass of the central system of Henize 2-428.
   The parameters of Henize 2-428 are in the region of VM,
   although we consider the errors.
   See text for more details.
}
 \label{fig12}
\end{figure}

\begin{figure}
 \begin{center}
   
        \includegraphics[width=8.4cm, angle=0]{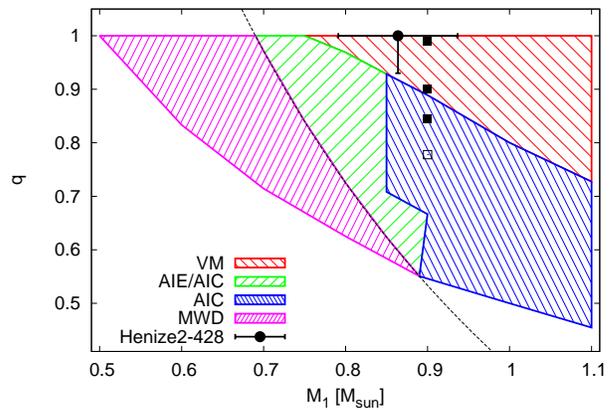}

 \end{center}
 \caption{Same as Figure \ref{fig13}, 
     but in the mass ratio versus primary mass diagram.
     Black squares denote the models calculated by \citet{pakmor11}.
     Filled squares indicate the models which satisfy 
     the detonation condition of \citet{seitenzahl09},
     while a open square does the model that does not.
}
 \label{fig13}
\end{figure}

We summarize the final fates of CO WD mergers in the total mass
versus mass ratio (Figure \ref{fig12}), as well as in the mass ratio
versus primary mass diagram (Figure \ref{fig13}).
The black dotted line shows the line of
$M_{\rm tot} = M_1 + M_2 = M_{\rm ig} = 1.38~M_\odot$.
If the total mass does not exceed this value,
the merger remnant becomes a WD without exploding
as an SN Ia \citep{nomoto84}.

The direct outcome is a violent merger-induced explosion 
(VM) in the red hatched region.
We adopt Equation (\ref{lower_bound}) as
the boundary of this immediate explosion.
In other regions, our condition for the violent merger scenario
is not satisfied and the merger remnant reaches 
a quasi-stationary state.

The merger remnant consists of three parts,
a cold core, hot envelope, and outer disk
\citep[e.g.][]{benz90}.
In the blue hatched region, off-center carbon burning occurs in
the hot envelope \citep[see, e.g.,][]{sato15} and the core of
the merger remnant would be converted to
an oxygen-neon-magnesium (ONeMg) WD
\citep{saio85,saio98,saio04}.
Because the total mass exceeds $M_{\rm ig}$,
the merger remnant finally collapses
to a neutron star, i.e., the accretion induced collapse (AIC)
\citep[e.g.,][]{kondo91}.

In the green hatched region,
the core of the merger remnant remains a CO WD because off-center
carbon burning does not occur in the early remnant phase
\citep[$10^2~-~10^3~{\rm s}$ after the secondary
is completely disrupted, e.g.,][]{sato15}.
Further evolution of the merger remnant depends
on the viscosity and the accretion of material
from the outer Keplerian disk \citep{yoon07}.
If the viscous heating is sufficiently large
in the viscous evolution phase
\citep[$10^4~-~10^8~{\rm s}$, e.g.,][]{mochkovitch89},
or if the accretion rate exceeds the critical rate of
${\sim}~2{\times}10^{-6}~M_{\odot}{\rm ~yr}^{-1}$
in the thermal evolution phase
\citep[$10^5q~-~10^6~{\rm yr}$, e.g.,][]{nomoto85, yoon07},
off-center carbon burning is ignited.
The final result would be AIC.
On the other hand, if the viscous heating is
too small or if the accretion rate does not
exceed the critical rate, the mass of
the merger remnant could grow to exceed
$M_{\rm ig}$ without off-center carbon burning.
We should note that $M_{\rm ig}$ of the rotating
CO WD is larger than $1.38~M_{\odot}$
because of the centrifugal force,
being $1.43~M_{\odot}$ if the rotation is uniform
(or larger if differential rotation).
Eventually, the merger remnant would
explode or collapse depending on $M_{\rm tot}$
and the timescale of the angular momentum
loss from the merger remnant \citep{benvenuto15}.
We call the case of explosion as the accretion
induced explosion (AIE).
Because of the above uncertainties,
we indicate the green hatched region as
AIE/AIC in Figure \ref{fig12}.
The boundary between the AIC and AIE/AIC is
derived from the results of \citet{sato15}.

In the magenta hatched region,
because off-center carbon burning
does not occur and the total mass of
the merger remnant does not exceed $M_{\rm ig}$,
it possibly leaves a massive CO WD (MWD).
Outer boundaries are set from the
WD mass range ($0.5~{\sim}~1.1~M_{\odot}$).

Figures \ref{fig12} and \ref{fig13} could not be conclusive.
If higher numerical resolutions are applied, 
the VM region in Figures \ref{fig12} and \ref{fig13}
would extend downward.
As mentioned in Section \ref{comparison},
the initial condition of simulations
also affects the merger outcomes.
If we adopt non-synchronously spinning systems
as initial conditions,
the VM and AIC region might extend
\citep[e.g.,][]{dan11, sato15}.

In Figure \ref{fig13}, we depict
the results of previous works for comparison.
The squares indicate the models calculated 
by \citet{pakmor11}.
Their WD masses are $M_2~=~0.70$, $0.76$, $0.81$, and $0.89~M_{\odot}$
for $M_1~=~0.9~M_{\odot}$ \citep{pakmor11}.
The filled squares are the models which satisfy the detonation
condition of \citet{seitenzahl09}, while the open one did not.
Their results indicate that $q_{\rm cr}~{\sim}~0.8$
for $M_1~=~0.9~M_{\odot}$.  Note that our boundary is 
$q_{\rm cr}~{\sim}~0.9$ for $M_1~=~0.9~M_{\odot}$, which
is more stringent than theirs.

We discuss the final fate of Henize 2-428 on
the basis of our results.
Henize 2-428 is a bipolar PN and
its central system was identified as a binary
\citep[e.g.,][]{rodriguez01}.
\citet{santander15} analyzed the light curve and
spectra of the central system.
From the depth of the light curve
minima, the intensities of two He II $541.2~$nm
absorption lines, and the obtained radial velocity amplitudes,
they interpreted the central system of Henize 2-428 as a DD
(double post-AGB core) system with nearly equal masses
and effective temperatures.
They estimated the total mass of the central system
as $1.76~M_{\odot}$, which clearly exceeds
the Chandrasekhar limit.
They also derived the merging time of this
system as ${\sim}~700~$Myrs from the orbital period of $0.176$~days.
If their interpretation is correct,
the central system of Henize 2-428 is the first detected
super-Chandrasekhar DD system,
which is a candidate of SN~Ia progenitors in the DD scenario.
Because the mass ratio is close
to unity, it likely leads to an SN~Ia
in the violent merger scenario.

\citet{santander15} fixed the mass ratio
of the central system of Henize 2-428
at unity based on the similar depths of
light curve minima and the radial velocity amplitudes.
In this paper, considering $1\sigma$-uncertainties mentioned in
\citet{santander15}, 
we derive possible ranges of the mass ratio and primary mass
without fixing the mass ratio.
In Figures \ref{fig12} and \ref{fig13}, the black point with error bars depicts
the possible ranges of the mass ratio and primary mass
thus derived.
Even considering the uncertainties of the mass ratio and primary mass,
we regard that Henize 2-428 is a possible progenitor
of SNe~Ia in the violent merger scenario.

\section{Summary}
\label{summary}
We summarize our main results as follows:

(1) Based on the SPH simulations of merging CO WDs,
we derived the critical mass ratio for
the violent merger scenario, i.e., 
the $q_{\rm cr}$ versus $M_1$ relation,
and compared our result with the previous studies.
Our $q_{\rm cr}(M_1)$ is more stringent than
that derived by \citet{pakmor11}.
We conclude that this small difference stems mainly
from the differences in the initial condition 
(synchronously spinning or not).

(2) We confirmed that the difference in the critical
mass ratio does not affect the brightness distribution of SNe Ia significantly,
as claimed in \citet{ruiter13}.
Our results are close to that of the ${\eta}~=~1.5$ case
in \citet{ruiter13} and consistent with the observational one obtained by
ROTSE-IIIb \citep{quimby12}.
Our larger $q_{\rm cr}$ would also decrease
the relative rate of higher central density primary WDs
at the merger, and as a result the total amount of $^{56}$Ni synthesized
in the violent merger would be reduced.

(3) We also summarized the direct outcome of CO WD mergers and
their final fates in the diagram of $M_{tot}$ versus $q$
(and $q$ versus $M_1$). 
On the basis of this diagram,
we discussed the fate of the central system of
the bipolar planetary nebula Henize 2-428, 
which was recently suggested to be a possible
super-Chandrasekhar DD system merging in a timescale much shorter than
the Hubble time.
Even considering uncertainties of the proposed system
in the $M_{tot}$ versus $q$ diagram,
we identify the final fate of this system as
almost certainly an SN~Ia in the violent merger scenario.

\section*{Acknowledgments}
We thank the anonymous referee for many detailed comments that helped
to improve the paper.
Simulations in this paper were performed by using computational resources of
Kavli Institute for the Physics and Mathematics of the Universe (IPMU),
and HA-PACS at the Center for Computational Sciences in University of
Tsukuba under Interdisciplinary Computational Science Program.
This research has been supported in part by Grants-in-Aid for Scientific
Research (23224004, 23540262, 23740141, 24540227, 26400222, and 26800100) 
from the Japan Society for the Promotion of Science and by the World Premier
International Research Center Initiative, MEXT, Japan.
This work is partly supported by MEXT program for the Development and Improvement 
for the Next Generation Ultra High-Speed Computer System under its Subsidies
for Operating the Specific Advanced Large Research Facilities.


\end{document}